\documentclass[reprint,apl,aip]{revtex4-1}

\usepackage{graphicx,epsfig}

\newcommand{\dC}{$^{\circ}$C}

\begin{document}
\title{Thermotropic Nematic and Smectic Order in Silica Glass Nanochannels}
\author{Andriy~V.~Kityk$^{1}$}
\email[E-mail: ]{andriy.kityk@univie.ac.at, p.huber@physik.uni-saarland.de}
\author{Patrick~Huber$^2$}

\affiliation{$^1$Faculty of Electrical Engineering, Czestochowa
University of Technology, Aleja Armii Krajowej 17, 42-200 Czestochowa,
Poland
\\$^2$Experimental Physics, Saarland University, D-66041 Saarbruecken, Germany
}

\date{\today}

\begin{abstract}
Optical birefringence measurements on a rod-like liquid crystal (8OCB), imbibed in silica channels (7 nm diameter), are presented and compared to the thermotropic bulk behavior. The orientational and positional order of the confined liquid evolves continuously at the paranematic-to-nematic and sizeably broadened at the nematic-to-smectic order transition, resp., in contrast to the discontinuous and well-defined second-order character of the bulk transitions. A Landau-de-Gennes analysis reveals identical strengths of the nematic and smectic ordering fields (imposed by the walls) and indicates that the smectic order is more affected by quenched disorder (originating in channel tortuosity and roughness) than the nematic transition.
\end{abstract}


\maketitle


The properties of spatially nano-confined liquid crystals (LCs) can be altered markedly compared to their bulk behavior \cite{Crawford, Bellini,Kutnjak,Kutnjak1, Guegan, Binder2008, Fish2010, Kityk1}. For example, high-resolution calorimetry revealed that the heat capacity anomaly in the vicinity of the second-order nematic-to-smectic-A (N-A) phase transition (PT), being evidently seen in the bulk, is
absent or at least significantly broadened upon confinement in aerogels \cite{Bellini}. Similarly, at temperatures $T$ even far above the bulk isotropic-to-nematic (I-N) PT there exists often a weak residual nematic ordering, a paranematic (PN) state \cite{Kralj}, which can be rigorously demonstrated by optical birefringence measurements performed on nematogen LCs confined in mesopores \cite{Kityk1}.

In this Letter we report an optical birefringence study on octyloxycyanobiphenyl (8OCB) confined in a mesoporous silica membrane permeated by parallel-aligned nanochannels (denotated as p-SiO$_{\rm 2}$). As we demonstrate here, it is  especially suitable for optical polarization studies both on the orientational and positional molecular ordering of nano-confined LCs. Moreover, its straight tubular pore topology renders p-SiO$_{\rm 2}$ particularly interesting for fundamental studies on spatially confined condensed matter systems aimed at a comparison of theory and experiment, as we shall demonstrate by a comparison of our results with a Landau-de-Gennes model for nematic and smectic order in cylindrical geometry.

\begin{figure}[htbp]
\epsfig{file=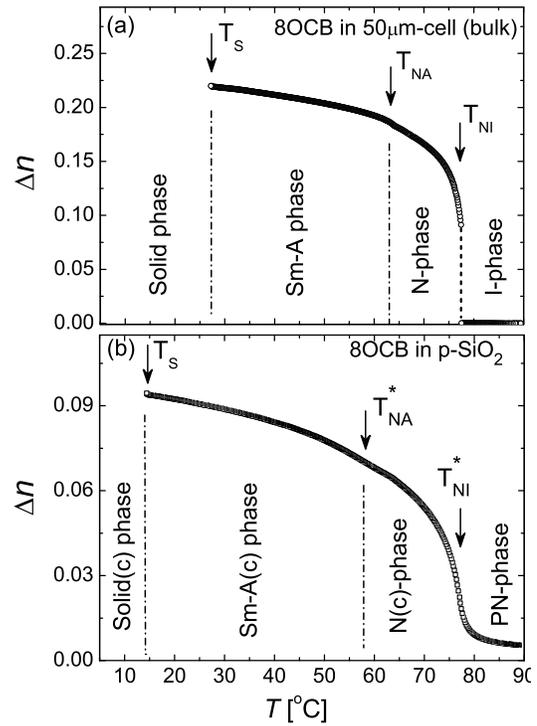, angle=0,
width=0.8\columnwidth}\caption{Measured optical birefringence $\Delta n$ vs. temperature $T$ for 8OCB in the bulk state (a) and confined in 7 nm pores of p-SiO$_{\rm 2}$ (b).} \label{fig1}
\end{figure}

p-SiO$_{\rm 2}$ has been prepared by electrochemical, anodic etching of highly doped $p$-type $<$100$>$ oriented silicon wafers. The membranes of $\sim$300 $\mu$m thickness have been subjected to thermal oxidation for 12 hours at $T=$800 $^o$C under standard atmosphere. The obtained monolithic SiO$_{\rm 2}$ membranes are permeated by channels perpendicular-aligned to the surface. The average pore diameter is 7~nm and the porosity is $\sim$ 48 \%, as determined by recording a N$_{\rm 2}$ sorption isotherm at $T$ = 77~K. The p-SiO$_{\rm 2}$ substrates were completely filled by spontaneous imbibition with 8OCB in the isotropic phase \cite{Huber2007, Gruener2010}. For the bulk measurements the sample cell, made of parallel glass plates ($d$=50 $\mu$m), has been filled by the LC in an homeotropically aligned state. The optical birefringence measurements were performed at a wavelength $\lambda$ of 632.8~nm on samples tilted out with regard to their optical axis (coinciding with the channel`s long axes) by 43 deg. The high-resolution optical polarization setup employs a photoelastic modulator and a dual lock-in detection scheme in order minimize the effects of uncontrolled light-intensity fluctuations \cite{Skarabot, Kityk1}. The accuracy of the optical retardation measurements is better than 5$\times$10$^{-3}$ deg. The optical birefringence calculated therefrom is a direct measure of the orientational (nematic) order parameter (OP) of the LC, the director. Moreover, it is also sensitive to the smectic OP, as we will discuss in more detail below.

\begin{figure}[htbp]
\vspace{-0.3cm}
\epsfig{file=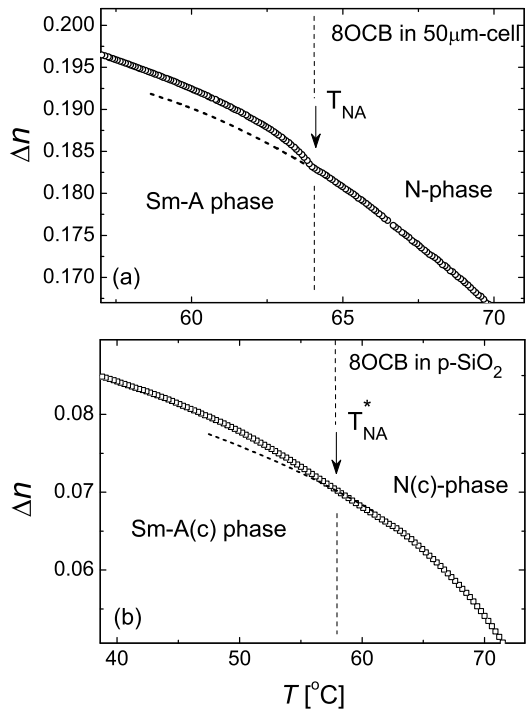, angle=0,width=0.8\columnwidth}
\vspace{-0.0cm}
\caption{Expanded plots of the temperature-dependent birefringence $\Delta n(T)$ in the regime of the nematic-to-smectic transition: (a) bulk, (b) confined in p-SiO$_{\rm 2}$.}
\label{fig2}\end{figure}

In Fig.~1 we show the optical birefringence of 8OCB measured upon slow cooling ($\sim$0.01 K/min) to the solidification temperature $T_{\rm S}$ in the bulk and the nano-confined states. Upon lowering $T$, bulk 8OCB exhibits three PTs: the I-N PT at $T_{\rm IN}\approx$77 $^o$C, the N-A PT at $T_{\rm NA}\approx$63.5 $^o$C and the smectic-A-to-crystalline PT at $T_{\rm S}\approx$27.0 $^o$C. The overall shape of the $\Delta n(T)$-dependence is in good agreement with earlier studies of Beaubois and Marcerou \cite{Beaubois}. However, due to the considerably higher resolution our measurements describe more precisely the characteristics of the N-A PT. It is continuous as one can infere from the expanded plot in Fig.~2 (a) and from an extrapolation (dashed line below $T_{\rm NA}$) of the much stronger increase in $\delta n$ typical of the orientational ordering gradually evolving below $T_{\rm IA}$. 

\begin{figure}[htbp]
\epsfig{file=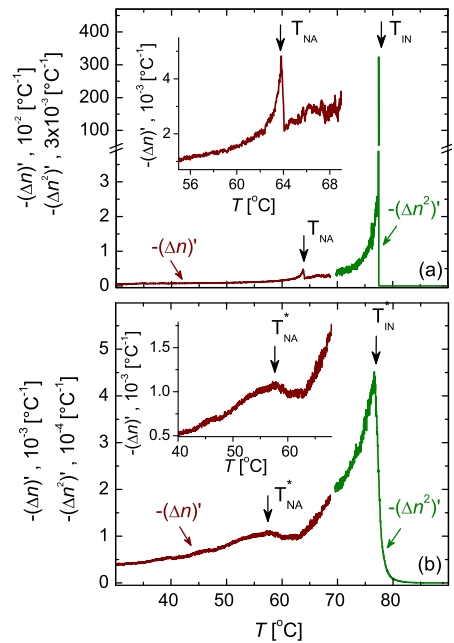, angle=0,width=0.7\columnwidth} \caption{Derivatives $(\Delta n)'$ and
$(\Delta n^2)'$ vs. temperature as determined from the data shown in Fig.~1. (a) bulk, (b) confined in p-SiO$_{\rm 2}$.} \label{fig3}
\end{figure}

The LC confined in p-SiO$_{\rm 2}$ exhibits a considerably different $T$-behavior, see Fig.~1 (b). The nematic ordering does not occur here via a discontinuous, first-order PT, but is clearly continuous in the vicinity of the bulk PT point $T_{\rm IN}$. Moreover, at temperatures even above $T_{\rm IN}$ there exists a weak residual nematic ordering characteristic of a paranematic state, documented by a specific pretransitional tail in the measured birefringence. Also the kink in the $\Delta n(T)$ dependence related with the second-order nematic-to-smectic-A PT, seen in bulk 8OCB, appears to be substantially smeared out upon confinement in p-SiO$_{\rm 2}$, see Fig.~2.

\begin{figure}[htbp]
\epsfig{file=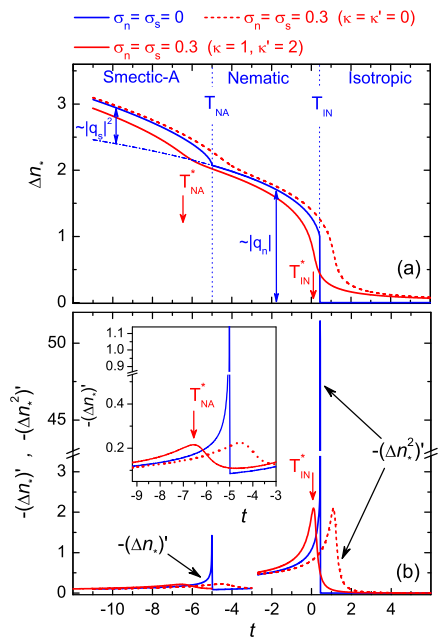, angle=0,width=0.7\columnwidth} \caption{Calculated normalized birefringence $\Delta n*$ vs. reduced temperature $t$ (a) and
$t$-derivatives $(\Delta n*)'$ and $(\Delta n*^2)'$ vs. $t$ (b) for selected surface field and quenched disorder parameters.} \label{fig4}
\end{figure}

A further description of the confinement effects requires the definition of the characteristic temperatures which separate the PN from the nematic (N) state ($T=T_{\rm IN}^*$) or the N and Smectic-A (Sm-A) states ($T=T_{\rm NA}^*$). One should be aware that both $T_{\rm IN}^*$ and $T_{\rm NA}^*$ have not any more a character of a PT in a classic meaning, since there is no spontaneous symmetry breaking in their vicinity. The full rotational symmetry is broken already by the anisotropic spatial confinement. Nevertheless, most of the physical properties keep anomalous behavior. As has been demonstrated in previous studies \cite{Kralj,Kutnjak,Kutnjak1,Kutnjak2} the peak position of the heat capacity anomaly, $C_{\rm p}(T)$, is a legitim indicator for a definition of the PT temperatures. Moreover, anomalous behavior is also exhibited by the $T$-derivative of the OP squared, $d\eta^2/dT$, being proportional to $-T^{-1}C_{\rm p}(T)$. Hence, taking into account that $\Delta n$ is proportional to the orientational (nematic) OP, $Q$, and is quadratic with respect to the positional (smectic) OP, $S$, the temperature derivatives $(\Delta n^2)'=d(\Delta n^2)/dT$ and $(\Delta n)'=d(\Delta n)/dT$ are suitable for an accurate determination of the PT temperatures.
In Fig.~3 we present the $T$-dependences of these derivatives extracted from the measured birefringence of 8OCB in the bulk and in the nanoconfined state. It is obvious that confinement leads to a considerable broadening of the derivative peaks, in accordance with previous calorimetry and diffraction studies on calamitic LCs confined in mesopores \cite{Kutnjak1,Kutnjak2,Guegan}.

In the following we analyze the obtained results with a Landau–de-Gennes model for the I-N and N-A PTs. The surface of amorphous silica pores enforce planar anchoring without a preferred in-plane direction \cite{Crawford}. However, the elongated geometry of cylindrical pores yields a preferred ordering along the pore axes. The dimensionless free energy density $f$ relevant to nematic ($f_{\rm n}$) and smectic ($f_{\rm s}$) ordering for such a geometry reads as \cite{Kutnjak1}: 
\begin{eqnarray}
f&=&f_{\rm n}+af_{\rm s}, \\
f_n&=&t_{\rm n} q_{\rm n}^2-2 q_{\rm n}^3+q_{\rm n}^4+c q_{\rm n}^6-q_{\rm n}\sigma_{\rm n}+\kappa_{\rm s} q_s^2,  \nonumber \\
f_{\rm s}&=& t_{\rm s} q_{\rm s}^2+\frac{1}{2} q_{\rm s}^4+c'q_{\rm s}^6-q_{\rm s}\sigma_{\rm s}+\kappa_{\rm n} q_n^2  \nonumber
\label{eq2}
\end{eqnarray}
where $q_{\rm n}$ and $q_{\rm s}$ are the normalized nematic and smectic order
parameters, $t_{\rm n}$ and $t_{\rm s}$ are nematic and smectic dimensionless reduced temperatures, $\sigma_{\rm n}$ and $\sigma_{\rm s}$ are the effective nematic and smectic ordering surface fields, respectively. A more detailed description of these models can be found in Ref. \cite{Kutnjak1}. 
The free energy expansion (1) is completed by $\kappa_{\rm n}$- and $\kappa_{\rm s}$-terms describing quenched disorder effects, originating from random orientational and positional field contributions, whereas the six order $c$- and $c'$-terms are included in order to reproduce the gradual saturation of the OP at low $T$. 

Minimization of the free energy (1) with respect to the OP $q_{\rm n}$ and $q_{\rm s}$ gives their equilibrium magnitudes $|q_{\rm n}|$ and $|q_{\rm s}|$. Due to the specific coupling between the optical electromagnetic wave, represented by the displacement vector $\vec{D}(\omega,k)$, and the OPs $q_{\rm n}$ and $q_{\rm s}$, which reads as $\vec{D_{\rm i}}(\omega,k)\vec{D_{\rm i}}^*(\omega,k)q_{\rm n}$ and $\vec{D_{\rm i}}(\omega,k)\vec{D_{\rm i}}^*(\omega,k)q_{\rm s}^2$, resp., the measured $\Delta n$ scales with $|q_{\rm n}|$ and $|q_{\rm s}|^2$. In Fig.~4 we show the normalized birefringence $\Delta n*=\Delta n(T)/\Delta n(T_{\rm IN})$ and its $t$-derivatives $(\Delta n*)'$ and $(\Delta n*^2)'$ vs. the reduced temperature $t$ being calculated for different sets of free energy parameters.
For vanishing surface fields (blue line) the model yiels discontinuous behavior and describes the overal behavior in very good agreement with our observation for the bulk system. In order to reproduce the continuous behavior in the pores, we had to use final, but identical values for the strength of the nematic and smectic ordering fields (see Fig.~3), and arrive thereby at a good description of all main features observed, including the PN ordering and broadening of the temperature peaks of the derivatives $(\Delta n)'$ and $(\Delta n^2)'$ in the vicinity of $T_{\rm IN}^*$ and $T_{\rm NA}^*$. Moreover, in order to account for the unshifted $T_{\rm IN}^*$ and significantly downshifted $T_{\rm NA}^*$ (by $\sim 6$ \dC), we have to use $\kappa_{\rm s}=2 \kappa_{\rm n}$=1 for the influence of quenched disorder on the smectic and nematic transition. Thus, a final tortuosity and pore wall roughness, presumably responsible for the quenched disorder contributions \cite{Crawford, Kityk1}, affect the layering formation significantly stronger than the evolution of orientational order \cite{FitNote}.  

Finally it is interesting to note that upon entering into the N-phase the bulk shear viscosity of 8OCB exhibits a minimum, resulting from an abrupt collective shear-alignment of the molecules. This rheological feature is, however, absent in silica mesopores, as was recently demonstrated by capillary filling experiments \cite{Gruener2010}. We believe the very gradual (and not abrupt) evolution of the nematic ordering upon nano-channel confinement, documented here, is responsible for this peculiarity. Thus, our study sheds not only important light on the equilibrium positional and orientational order of an archetypical rod-like LC, it allows also for a better understanding of the non-equilibrium behavior of this system upon nano-confinement. 


\begin{thebibliography}{00}
%
\bibitem{Crawford} G.P. Crawford and S. Zumer (Eds.), Liquid Crystals in Complex Geometries, Taylor \& Francis (1996).
\bibitem{Bellini} T. Bellini, L. Radzihovsky, J. Toner, and N. A. Clark, Science \textbf  {294}, 1074 (2001).






\bibitem{Kutnjak} Z.~Kutnjak, S.~Kralj, G.~Lahajnar, and S.~Zumer, Phys. Rev. E \textbf{68}, 021705 (2003).

\bibitem{Kutnjak1} Z.~Kutnjak, S.~Kralj, G.~Lahajnar, and S.~Zumer, Phys. Rev. E \textbf{70}, 051703 (2004).
\bibitem{Guegan}  R. Guegan, D. Morineau, C. Loverdo, W. Beziel, and M. Guendouz, Phys. Rev. E \textbf{73}, 011707 (2006).
\bibitem{Binder2008} K. Binder, J. Horbach, R. Vink, and A. de Virgiliis, Soft Matter \textbf{4}, 1555 (2008).
\bibitem{Fish2010} J. M. Fish and R. L. C. Vink, Phys. Rev. Lett. \textbf{105}, 147801 (2010).
\bibitem{Kityk1} A.~V. Kityk, M.~Wolff, K.~Knorr, D.~Morineau, R.~Lefort, and P.~Huber, Phys. Rev. Lett. \textbf{101}, 187801 (2008).

\bibitem{Kutnjak2} Z.~Kutnjak, S.~Kralj, and S.~Zumer, Phys. Rev. E \textbf{66}, 041702 (2002).
\bibitem{Huber2007} P. Huber, S. Gr\"uner, C. Sch\"afer, K. Knorr, and A. V. Kityk, Europ. Phys. J. Spec. Top. \textbf{141}, 101 (2007).
\bibitem{Gruener2010} S. Gruener and P. Huber, arXiv:1005.0730. 
\bibitem{Skarabot} M.~Skarabot, M.~Cepic, B.~Zeks, R.~Blinc, G.~Heppke, A.~V.~Kityk, and I. Musevic, Phys. Rev. E \textbf{58} 575 (1998).
\bibitem{Beaubois} F.~Beabois and J.~P.~Marcerou, Europhys. Lett. \textbf{36 } 111 (1996).
\bibitem{FitNote} Note that a full-fitting of the experimental results with the model is hampered by the neglection of higher order terms in Eq.~1 necessary to describe the correct saturation of the OP.
\end{thebibliography}
\end{document}